# Some basic features of canonical ensemble in noncommutative spaces

S. A. Alavi

Department of Physics, Hakim Sabzevari University, P. O. Box 397, Sabzevar, Iran

s.alavi@hsu.ac.ir ; alaviag@gmail

**Abstract**. We calculate the corrections due to noncommutativity of space on the Hamiltonian and then partition function of the canonical ensemble. We study some basic features of statistical mechanics and thermodynamics including equipartition and virial theorem and energy fluctuations: correspondence with microcanonical ensemble, in the framework of non-commutative canonical ensemble. The corrections imposed by noncommutativity of space are derived and the results are discussed.

## 1. Introduction

Statistical Mechanics is one of the pillars of modern physics and is a valuable tool, which investigates the macroscopic nature of physical systems in terms of dynamical laws controlling the microscopic constituents of the systems. On the other hand, noncommutative geometry is a very general mathematical pattern which extends throughout different branches of mathematics and physics. Heisenberg suggested the concept of noncommutativity of space as a possible strategy to remove the divergences in quantum field theory. In 1999, it was shown by some researches that [1,2] the noncommutative coordinates appear consistently in string theory, which generated an immense study of the subject since then. Over the last few years, there have been significant attempts to formulate physical theories in noncommutative spaces, ranging from condensed matter to high energy physics, astrophysics and cosmology. Noncommutative geometry is also a well-known framework to implement quantum gravity effects in a physical system. The noncommutative coordinates satisfy the following commutation relation:

$$[\hat{x}_\mu, \hat{x}_\nu] = i\theta_{\mu\nu} \tag{1}$$

where $\theta_{\mu\nu}$ is the noncommutativity parameter. Using Bopp-shift [3], one can relate the $\theta$-variables to the conventional commutative space variables:

$$\hat{x}_i = x_i - \frac{1}{2\hbar}\theta_{ij}p_j, \qquad \hat{p}_i = p_i, \quad \text{So} \quad \frac{\partial}{\partial \hat{x}_i} = \frac{\partial}{\partial x_i} \tag{2}$$

So the Hamiltonian in NC space might be written as

$$H^\theta(\hat{x}_i, \hat{p}_i) = H(x_i - \frac{1}{2\hbar}\theta_{ik}p_k, p_i) = H(x_i, p_i) + \Delta H^\theta(x_i, p_i) \tag{3}$$

where

$$\Delta H^\theta(x_i, p_i) = -\frac{1}{2\hbar}\theta_{\ell k}p_k \frac{\partial H}{\partial x_\ell} \tag{4}$$

## 2. The equipartition theorem

The equipartition theorem states that energy is shared equally amongst all energetically accessible degrees of freedom of a system. To study this theorem in noncommutative spaces let's determine the expectation value of the quantity

$\langle \hat{q}_i \frac{\partial H^\theta}{\partial \hat{q}_j} \rangle$ , where $q$ is any of the $6N$ generalized coordinates $x$ and $p$.

**Case 1:** $\hat{q}_i = \hat{x}_i \qquad \hat{q}_j = \hat{p}_j$

$$\langle \hat{x}_i \frac{\partial H^\theta}{\partial \hat{p}_j} \rangle = \langle (x_i - \frac{1}{2\hbar}\theta_{ik}p_k)\frac{\partial}{\partial p_j}(H + \Delta H^\theta) \rangle =$$

$$\langle x_i \frac{\partial H}{\partial p_j} \rangle - \frac{\theta_{\ell k}}{2\hbar}\langle x_i \frac{\partial}{\partial p_j}(p_k \frac{\partial H}{\partial x_\ell}) \rangle - \frac{\theta_{ik}}{2\hbar}\langle p_k \frac{\partial H}{\partial p_j} \rangle + O(\theta^2) \tag{5}$$

But

$$\langle x_i \frac{\partial H}{\partial p_j} \rangle = 0, \qquad \langle p_k \frac{\partial H}{\partial p_j} \rangle = \langle x_k \frac{\partial H}{\partial x_j} \rangle = \delta_{kj}kT$$

and

$$-\frac{\theta_{\ell k}}{2\hbar}\langle x_i \frac{\partial}{\partial p_j}(p_k \frac{\partial H}{\partial x_\ell}) \rangle = -\frac{\theta_{\ell k}}{2\hbar}\langle x_i \frac{\partial H}{\partial x_\ell} \rangle \delta_{kj} - \frac{\theta_{\ell k}}{2\hbar}\langle x_i p_k \frac{\partial^2 H}{\partial p_j \partial x_\ell} \rangle$$

Hence:

$$\langle \hat{x}_i \frac{\partial H^\theta}{\partial \hat{p}_j} \rangle = -\frac{1}{\hbar}\theta_{ij}kT - \frac{\theta_{\ell k}}{2\hbar}\langle x_i p_k \frac{\partial^2 H}{\partial p_j \partial x_\ell} \rangle \tag{6}$$

For an ideal gas with the Hamiltonian $H = \sum_{i=1}^{3N} \frac{p_i^2}{2m}$ which N is the number of the particles, the second term vanishes. It is interesting to have an estimate of the magnitude of these corrections. As stated above, for an ideal gas the second term in (6) vanishes. About first term for a monoatomic gas is room temperature i.e., T=300 K, the average kinetic energy of a gas molecule is:

$$\langle H \rangle = \langle \sum_{i=1}^{3} \frac{p_i^2}{2m} \rangle = \frac{3}{2} kT = \frac{3}{2} \times 1.38 \times 10^{-23} \times 300 = (0.038 ev) \tag{7}$$

Furthermore, using the upper bound on the noncommutativity parameter $\theta = (10^4 GeV)^{-2}$ [3,4], one can estimate the value of the correction term $\frac{1}{\hbar}\theta kT$, as $6 \times 10^{-14} \frac{1}{s\,ev^2}$ which shows that the correction is negligibly small as expected. By increasing the energy of the constituent particles of the system, the effects of noncommutativity of space increase.

**Case 2:** $\hat{q}_i = \hat{p}_i \quad \hat{q}_j = \hat{p}_j$

$$\langle \hat{p}_i \frac{\partial H^\theta}{\partial \hat{p}_j} \rangle = \langle p_i \frac{\partial H^\theta}{\partial p_j} \rangle = \delta_{ij} kT - \frac{\theta_{\ell k}}{2\hbar} \langle p_i p_k \frac{\partial^2 H}{\partial p_j \partial x_\ell} \rangle \tag{8}$$

The second term has a nonzero value in general but vanishes for the case of an ideal gas.

**Case 3:** $\hat{q}_i = \hat{x}_i \quad \hat{q}_j = \hat{x}_j$

$$\langle \hat{x}_i \frac{\partial H^\theta}{\partial \hat{x}_j} \rangle = \langle (x_i - \frac{1}{2\hbar}\theta_{ij} p_j) \frac{\partial H^\theta}{\partial x_j} \rangle = \langle (x_i - \frac{1}{2\hbar}\theta_{ij} p_j) \frac{\partial (H - \frac{1}{2\hbar}\theta_{\ell k} p_k \frac{\partial H}{\partial x_\ell})}{\partial x_j} \rangle \tag{9}$$

Therefore:

$$\langle \hat{x}_i \frac{\partial H^\theta}{\partial \hat{x}_j} \rangle = \langle x_i \frac{\partial H}{\partial x_j} \rangle - \frac{1}{2\hbar}\theta_{\ell k} \langle x_i p_k \frac{\partial^2 H}{\partial x_j \partial x_\ell} \rangle - \frac{1}{2\hbar}\theta_{ij} \langle p_j \frac{\partial H}{\partial x_j} \rangle + O(\theta^2)$$

The second term is an odd function on the momentum, so its average gives zero contribution. The average in the third term also vanishes. So, we have:

$$\langle \hat{x}_i \frac{\partial H^\theta}{\partial \hat{x}_j} \rangle = \langle x_i \frac{\partial H}{\partial x_j} \rangle = \delta_{ij} kT \tag{10}$$

**Case 4:** $\hat{q}_i = \hat{p}_i \quad \hat{q}_j = \hat{x}_j$

$$\langle \hat{p}_i \frac{\partial H^\theta}{\partial \hat{x}_j} \rangle = \langle p_i \frac{\partial H^\theta}{\partial x_j} \rangle = -\frac{\theta_{\ell k}}{2\hbar} \langle p_i p_k \frac{\partial^2 H}{\partial x_j \partial x_\ell} \rangle \tag{11}$$

Comparison of cases 1 to 4 shows that the equipartition theorem is not valid in NC spaces.

## 3. Energy fluctuations in the canonical ensemble: Equivalence of the microcanonical and the canonical ensembles in the thermodynamic Limit

In commutative spaces, it is a common technique that by calculation of energy fluctuation in canonical ensemble as appears below, it is argued that microcanonical and canonical ensembles behave in exactly the same way in the thermodynamic limit. The energy fluctuation in canonical ensemble is given by [5]:

$$\langle (\Delta E)^2 \rangle = \langle E^2 \rangle - \langle E \rangle^2 = -\frac{\partial U}{\partial \beta} = kT^2 C_v \tag{12}$$

Where $C_v$ is the specific heat at constant volume. For the relative root-mean-square fluctuation in $E$ we have:

$$\frac{\sqrt{\langle (\Delta E)^2 \rangle}}{U} = \frac{\sqrt{kT^2 C_v}}{U} \tag{13}$$

which is of the order of $(N^{-\frac{1}{2}})$. Therefore, for large $N$ (which is the case for every statistical system) the relative r.m.s. fluctuation in the values of $E$ is fully negligible. Thus, for all practical purposes, a system in the canonical ensemble has an energy equal to, or almost equal to, the mean energy $U$. So the situation is practically the same as in the microcanonical ensemble. To examine this point in noncommutative spaces, we start with the expression for the mean energy

$$U_\theta = \langle E_\theta \rangle = \frac{\sum_r E_r^\theta \exp(-\beta E_r^\theta)}{\sum_r \exp(-\beta E_r^\theta)} \tag{14}$$

But we can write $E_r^\theta = E_r + \Delta E_r^\theta$, where $E_r$ and $\Delta E_r^\theta$ are the energy of the system in commutative space and the correction due to noncommutativity of space respectively. To the first order in θ, $\Delta E_r^\theta = \theta \Delta E_r$. After some calculations to the first order in θ we have:

$$\langle E \rangle_\theta = U_\theta = U - \beta\theta\langle E \Delta E \rangle + \theta\langle \Delta E^\theta \rangle \tag{15}$$

U is the mean energy of the system in commutative space. One can also drive that:

$$\langle E^2 \rangle_\theta = \langle E^2 \rangle - \beta\theta \langle E^2 \Delta E^\theta \rangle \tag{16}$$

Therefore, we achieve the following expression for the energy fluctuations:

$$\langle (\Delta E_\theta)^2 \rangle = \langle (\Delta E)^2 [1 - \beta\theta\Delta E^\theta] \rangle + \beta\theta U^2 \Delta E^\theta \tag{17}$$

where $\langle (\Delta E_\theta)^2 \rangle$ and $\langle (\Delta E)^2 \rangle$ are the energy fluctuation of the system in noncommutative and commutative spaces respectively. Here $\theta \Delta E^\theta = \theta \langle \Delta E_r^\theta \rangle$ is the ensemble average of the energy correction caused by noncommutativity of space on the system. For the relative root-mean-square fluctuation in $E$, we obtain:

$$\frac{\sqrt{\langle (\Delta E_\theta)^2 \rangle}}{U} = \frac{\sqrt{\langle (\Delta E [1 - \beta\theta \Delta E_\theta])^2 \rangle}}{U} + \sqrt{\beta\theta \Delta E_\theta} \tag{18}$$

The second term in r.h.s is independent of the number of particles and the size of the system. So it does not vanish even for the systems with large number of particles which is in contrast to the commutative case. For that reason, the correspondence between canonical and microcanonical ensembles is violated in noncommutative spaces.

**4. The virial theorem in NC space**

The virial theorem is *a relation between the total kinetic energy and the total potential energy of a system in* equilibrium. The definition and the value of the virial of a system in commutative spaces are as follows:

$$V_0 = \langle x_i F_i \rangle = -\langle x_i \frac{\partial H}{\partial x_j} \rangle = -3NkT \tag{19}$$

The virial of the system in NC space is

$$V_\theta = -\langle \hat{x}_i \frac{\partial H^\theta}{\partial \hat{x}_j} \rangle = \langle \hat{x}_i \dot{\hat{p}}_i \rangle = \langle (x_i - \frac{1}{2\hbar}\theta_{ik} p_k) \dot{p}_i \rangle = \langle x_i \dot{p}_i \rangle - \frac{1}{2\hbar}\theta_{ik} \langle p_k \dot{p}_i \rangle \tag{20}$$

It could be simplified as:

$$V_\theta = \langle x_i \dot{p}_i \rangle - \frac{1}{2\hbar}\theta_{ik} \langle p_k \dot{p}_i \rangle = \langle x_i F_i \rangle - \frac{1}{2\hbar}\theta_{ik} \langle p_k F_i \rangle = V_0 - \frac{1}{2\hbar}\theta_{ik} \langle p_k F_i \rangle \tag{21}$$

$V_0$, is the virial of the system in commutative space. In the case of a classical gas consisting of particles with no interactions, the only forces that come into operation are the ones arising from the walls of the container. If we take P as the pressure of the system, Then the force corresponding to an element of area dσ of the walls is –Pdσ. The negative sign appears because the force is directed inward while the vector dσ is directed outward. So for the second term we have:

$$-\frac{1}{2\hbar}\theta_{ik}\langle p_k F_i\rangle = -\frac{1}{2\hbar}(\frac{1}{2}\varepsilon_{ikj}\theta_j)\langle p_k F_i\rangle = \frac{1}{4\hbar}\theta_j\langle(\vec{p}\times\vec{F})_j\rangle = \frac{1}{4\hbar}P\theta_j\oint(\vec{p}\times\vec{d\sigma})_j \qquad (22)$$

where we have used $\theta_{ij} = \frac{1}{2}\varepsilon_{ijk}\theta_k$ [3,4]. Using an alternative form of Gauss theorem:

$$\oint \vec{p}\times\vec{d\sigma} = \int(\nabla\times\vec{p})d\tau \qquad (23)$$

It could be written as follows

$$\frac{1}{4\hbar}P\theta_j\int(\nabla\times\vec{p})_j d\tau = 0 \qquad (24)$$

Because, for this system the curl of the momentum is zero. So correction due to noncommutativity of space on virial theorem for an ideal gas is zero but it does not vanish for other systems in general.

5. Conclusions

We have shown that:

● The equipartition theorem is not valid in noncommutative spaces.
● The canonical and microcanonical ensembles are not equivalent in the thermodynamics limit in noncommutative spaces.
● The virial theorem remains valid in noncommutative spaces for ideal systems for which the only forces which come into play are the one arising from the walls of the container.